\documentclass[12pt]{iopart}
\usepackage{times}
\usepackage{epsfig}
\usepackage{graphicx}
\usepackage{dcolumn}
\usepackage{natbib}

\def\beginwide{
        \end{multicols} \vspace*{-0.5cm} \noindent
        \rule{3.5in}{.1mm}\rule{.1mm}{5mm} \widetext \medskip }
\def\beginwidetop{
        \end{multicols} \vspace*{-0.5cm} \noindent
        \widetext \medskip }
\def\endwide{
        \hspace*{3.35in}~\rule[-5mm]{.1mm}{5mm}\rule{3.5in}{.1mm}
        \begin{multicols}{2} \vspace*{-1.0cm} \noindent }
\def\endwidebottom{
        \begin{multicols}{2} \vspace*{-1.0cm} \noindent }

\begin{document}
\title{Morphology of two dimensional fracture surfaces}
\author{Mikko J. Alava$^1$, Phani K. V. V. Nukala $^2$ and Stefano Zapperi$^3$}
\address{$^1$ Helsinki University of Technology, Laboratory of Physics,  HUT-02105 Finland}
\address{$^2$ Computer Science and Mathematics Division, 
Oak Ridge National Laboratory, Oak Ridge, TN 37831-6164, USA   }
\address{$^3$ CNR-INFM, SMC, Dipartimento di Fisica,
Universit\`a "La Sapienza", P.le A. Moro 2
00185 Roma, Italy}

\begin{abstract}
We consider the morphology of two dimensional cracks observed in 
experimental results obtained from paper samples and compare these results 
with the numerical
simulations of the random fuse model (RFM). We demonstrate that
the data obey multiscaling at small scales but cross over
to self-affine scaling at larger scales. Next, we show that
the roughness exponent of the random fuse model is recovered by a simpler model 
that produces a connected crack, while a directed crack 
yields a different result, close to a random walk.
We discuss the multiscaling behavior 
of all these models.
\end{abstract}
\maketitle

Understanding the statistical properties of fracture surfaces
has been an important theoretical challenge for the past twenty
years, starting from the first pioneering experimental 
evidence of self-affinity provided by Mandelbrot et al. \cite{mandelbrot84}. 
Most experimental results 
reported in the past for three dimensional fracture surfaces 
suggested the presence of a universal roughness 
exponent in the range  $\zeta \simeq 0.75-0.85$ \cite{bouchaud97}.
The scaling regime is sometimes quite impressive,
spanning five decades in metallic alloys \cite{bouchaud97}.
Here, it was argued  that a different exponent 
(i.e. $\zeta \simeq 0.4-0.5$) should describe the small scales, 
with a crossover originally interpreted 
as a dynamic effect. This exponent would correspond to 
the quasistatic limit while the large scale exponent to the effect of finite
velocities \cite{bouchaud97}. It was recently pointed out that the short-scale 
value is not present in silica glass, even when cracks
move at extremely low velocities \cite{ponson06}. In addition,
in granite and sandstone, one only observes the "small scale" 
exponent even at high velocities  \cite{boffa98}. The current
interpretation associates
the value $\zeta\simeq 0.75$ with rupture processes occurring
inside the fracture process zone, where elastic interactions
would be screened and the value $\zeta \simeq 0.45$ with large
scale elastic fracture \cite{bonamy06}. The authors of 
Refs.~ \cite{ponson06,bonamy06} were also able to show that
the fracture surface is anisotropic, with different exponents in 
parallel and perpendicular directions to the crack propagation. 
In addition, we have to remark that the measured roughness exponent
describes only the local properties of the surface. 
The fracture surface in many cases 
exhibits anomalous scaling: the {\it global} exponent describing the
scaling of the crack width with the sample size is larger than
the local exponent measured on a single sample
(refs.~\cite{lopez97,morel98,mourot05}). It is thus necessary to
define two roughness exponents a global ($\zeta$) and a local $\zeta_{loc}$.

Two dimensional fracture surfaces are in principle
simpler to analyze than the three-dimensional surfaces 
which can be anisotropic
and since the crack surface reduces to
a line in 2D. The existing experimental results, obtained mainly in paper,
point towards a (local)  roughness exponent in the range $\zeta \simeq 0.6-0.7$ 
\cite{kertesz93,engoy94,salminen03,rosti01}. 
However, even for ordinary, industrial paper itself there are numerous values available 
that are significantly higher than $\zeta=0.7$ (examples are found in Ref.~\cite{menezessobrinho05}). 
It is not known at this time whether
this variation in $\zeta$ values is a reflection of 
difficulty in experimentally measuring $\zeta$,
or that it is not really universal but depends on
material parameters such as ductility and anisotropy. 
Recently, Bouchbinder et al. have indicated a scenario in which the
the crack line $h(x)$ has more complicated structure, exhibiting {\em
multiscaling} behavior. This implies a non-constant
exponent $\alpha_q$, for the $q$-th order
correlation function $C_q(x) = \langle |h(x+y)-h(y)|^q\rangle^{1/q}
\sim x^{\alpha_q}$,
\cite{bouchbinder05a}. This result would strongly put into 
question the existence of a well defined roughness exponent in two 
dimensional fracture. It should also be noted that so far there is no
experimental evidence for the
presence of anomalous scaling in two dimensions.

From the theoretical point of view, two dimensional fracture 
could appear as a relatively simple problem if we consider
the crack surface as the trace left by a point (i.e. the crack tip) 
moving through a disordered elastic medium \cite{bouchaud93,daguier97}. 
A similar idealization
is also used in three dimensions where the crack tip is replaced
by a deforming crack line front. Under mode I quasistatic loading,
the fracture surface is shown to be only logarithmically rough in three dimensions,
in contrast with the experiments \cite{ramanathan97}.
This suggests that to understand the experiments one should consider 
additional ingredients, such as damage nucleation ahead of the tip \cite{celarie03}, 
crack branching or elastodynamics effects. 
In this perspective, disordered lattice models
provide an alternative way to describe the phenomenon \cite{herrmann90,alava06}. 
In these the elastic medium is described by a network of 
springs with random failure thresholds. In the simplest approximation of a 
scalar displacement, one recovers the random fuse model (RFM) where a lattice of fuses with
random threshold are subject to an increasing external voltage
\cite{dearcangelis85,kahng88}.
The model has been numerically simulated to obtain the roughness of 
the fracture surface in two \cite{hansen91b,raisanen98,seppala00,zapperi05} 
and three dimensions \cite{batrouni98,raisanen98,raisanen98b,nukala_fuse3d}.
In two dimensions, the roughness exponent $\zeta_{loc}\simeq 0.7$ is 
reasonably close to the experimental results. In addition,
recent simulations reveal that anomalous scaling is also present,
although the effect is small (i.e. $\zeta-\zeta_{loc}=0.1$)
\cite{zapperi05}. 

In this letter we first point out that the multiscaling behavior observed
in Ref.~\cite{bouchbinder05a} is in fact a small scale effect due to
the fluctuations induced by the fibrous structure of paper. To
this end we re-analyze cracks produced in 6500 $mm$ long paper samples \cite{salminen03}. 
We next turn our attention to the RFM and observe
that similar corrections exist there as well. In order 
to understand the mechanism underlying the roughness in the RFM
we introduce two simple models that describe the
growth of a single connected crack. The numerical results show
that a fracture process zone (FPZ) is essential in recovering the RFM scaling.
We thus conclude that considering the motion of a crack tip
represents an oversimplification of the problem, leading to
quantitatively different results.

\begin{figure}[hbtp]
\begin{center}
\includegraphics[height=6cm]{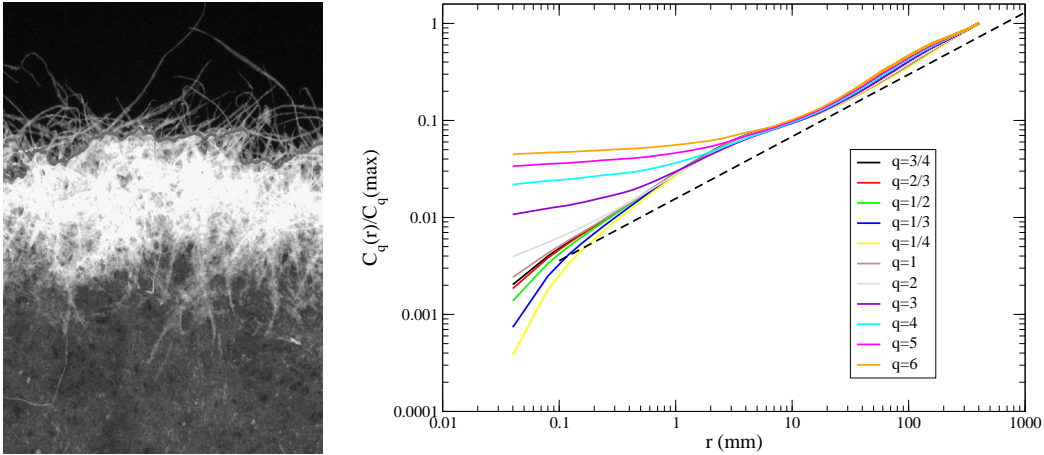}
\hspace{0.3cm}
\includegraphics[height=6cm]{corr2.eps}
\end{center}
 \caption {The diffuse crack surface in paper
and the thresholded ``interface''. 
At small scale it is not obvious how
to define a single valued interface and jumps associated
with overhangs are unavoidable (left). As a result of this
one sees an apparent multiscaling at small scales, followed
by self affine scaling at larger scales. The curves represent
the $q-$correlation functions, normalized by their maximum
value for clarity (right). The dashed line has an exponent $\zeta=0.64$.}
\label{fig:1}
\end{figure}

Paper would seem to be a good test material for fracture
surface analysis since it has a disordered structure and
can be considered effectively two-dimensinal (for an in-depth
discussion see Ref. \cite{alava06b}). However, as in any similar
problem, the coarse-grained behavior can only be seen at scales
that are larger than those associated with the microscopic details.
Fig. \ref{fig:1} illustrates (left panel) the intricacies of 
fracture line analysis in paper. First of all, on a scale below
(typically) 0.1 $mm$, paper is in fact three-dimensional and 
the fracture surface is no longer a line. Second, the damage is
diffusive on the typical FPZ scale, which ranges in industrial
papers up to 2-3 $mm$. The greyscale in the figure shows in 
fact such a damage profile and illustrates that on such
scales the final single-valued fracture line will have steep
gradients $\Delta h$. On a scale $\Delta x$ of the order of the microstructural details
the distribution $P (\Delta h) (\Delta x)$ displays non-Gaussian tails
\cite{salminen03}. The $C_q (x)$ can be sensitive to such tails, when
the height differences $|h(x+x')-h(x')|$ that dominate
$C_q$ are also important for the $q$ moment of $P(\Delta h)$.
This is demonstrated in the right-hand panel of Fig.~\ref{fig:1}, 
where we plot the  $C_q$ for various moments $q$
ranging from $q=1/4$ to $6$. It is clear that there is apparent
multiscaling related to the $P(\Delta h)$ (see \cite{salminen03})
and that beyond a cross-over scale of a few $mm$ one obtains
self-affine scaling with $\zeta = 0.64$ in agreement with
other measures \cite{salminen03}. In passing, we note that the structure
of paper exhibits correlations (due to so-called flocs) over
still larger scales than the FPZ size extending up to several $mm$. 
The cross-over scale might be related to these correlations.

In the RFM \cite{dearcangelis85}, one considers a triangular lattice 
with fuses having all the same conductance and
random breaking thresholds $t$, uniformly distributed between 0 and
1. The burning of a fuse occurs irreversibly, whenever the electrical
current in the fuse exceeds its threshold $t$. 
Periodic boundary conditions are imposed in the horizontal
direction and a constant voltage
difference, $V$, is applied between the top and the bottom of lattice
system bus bars. Numerically, a unit voltage difference, $V = 1$, is
set between the bus bars and the Kirchhoff equations are solved to
determine the currents in the fuses. Subsequently, for
each fuse $j$, the ratio between the current $i_j$ and the breaking
threshold $t_j$ is evaluated, and the fuse $j_c$ having the largest
value, $\mbox{max}_j \frac{i_j}{t_j}$, is irreversibly removed
(burnt).  The current is redistributed instantaneously after a fuse is
burnt implying that the current relaxation in the lattice system is
much faster than the breaking of a fuse.  Each time a fuse is burnt,
it is necessary to re-calculate the current redistribution in the
lattice to determine the subsequent breaking of a fuse.  The process
of breaking of a fuse, one at a time, is repeated until the lattice
system fails completely, producing an irregular fracture surface.

Here, we propose two variations to the RFM. (i) In the first variation, 
we impose that failure events form a connected crack, excluding
damage nucleation in the bulk. This means that after breaking the weakest fuse,
successive failure events are only allowed on fuses that are connected
to the crack. Otherwise, the rules of this simplified model strictly follow
those of the usual RFM. In effect this rule implies that
we have a FPZ which is constrained to a distance $r=1$ from an evolving 
crack. (ii) In the second variation, in addition to the variation (i), 
we also do not allow for any crack branching, nor turning backwards.
We thus only break one of the three fuses connected to the crack tips 
(i.e. the one with the largest current/threshold ratio).
This leads to a single directed crack, with the surface
being the trail left by the crack tip.

We simulate the variations (i) and (ii) on triangular
lattices of linear sizes $L=128,192,256,320,512$ with 
uniformly distributed disorder.  The final crack
in case (i) and in the RFM typically displays some limited amount of dangling 
ends and overhangs. 
These are removed to obtain a single valued crack line $h(x)$. 
This is not necessary for case (ii) since the crack is by definition
single valued in this scenario. 
Several methods have been devised to characterize the roughness of
a self-affine interface. 
The power spectrum is believed to provide one of the most reliable 
estimates of the roughness exponent, and decays as
\begin{equation}
S(k)\equiv  \sum_x e^{i(2\pi xk/L)} \langle h(x)h(0) \rangle 
\sim k^{-(2\zeta+1)}.
\end{equation}
It was shown in Ref.~\cite{zapperi05} that the RFM 
displays anomalous scaling, and in this case, the power spectrum scales
as $S(k,L) \sim k^{-(2\zeta_{loc}+1)}L^{2(\zeta-\zeta_{loc})}$.

\begin{figure}[hbtp]
\begin{center}
\includegraphics[width=8cm]{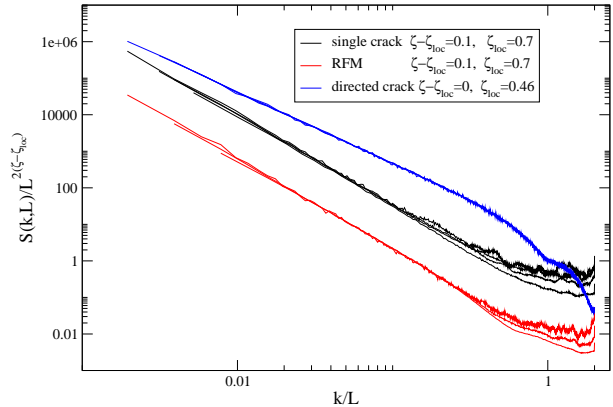}
\end{center}
\caption{The power spectrum of the crack $S(k,L)$ obtained from different models for different lattice sizes in log-log scale.  
The slope defines the local exponent as $-(2\zeta_{loc}+1)$. The spectra for all of the different lattice sizes can be collapsed 
using simple self-affine scaling only for directed cracks,
yielding $\zeta=\zeta_{loc}=0.46$. Both the RFM 
and the single crack model show instead anomalous scaling
with $\zeta_{loc}=0.7$ and $\zeta=0.8$}
\label{fig:2}
\end{figure}

In Fig.~\ref{fig:2} we compare the power spectra of the cracks
obtained from the RFM \cite{zapperi05} and from variations (i) and (ii).
The RFM and variation (i) follow essentially the same scaling
behavior with $\zeta_{loc}\simeq 0.7$ and $\zeta\simeq 0.8$.
On the other hand, the directed crack scenario (variation (ii)) 
differs considerably: there is no anomalous scaling and the
exponent is significantly lower (i.e. $\zeta=\zeta_{loc}\simeq 0.46$).
This result is very close to the random walk exponent
$\zeta=1/2$ that would be found in the directed crack model
wherein the effects of current are completely ignored. 
In this scenario, the crack tip would move up and down depending only on 
the smaller threshold. The fact that the exponent is slightly
smaller than $\zeta=1/2$ could be a numerical artifact associated 
with random walk on triangular lattice topology. 
Hence, to recover the RFM roughness exponents, the presence of
crack branching, as in variation (i), is essential.  
This result indicates that the presence of distant
damage nucleation and coalescence with other cracks
is irrelevant for the crack surface roughness. 

In Fig.~3, we analyze the multscaling behavior of the
RFM and its two variations. As expected, the directed crack 
does not display any multiscaling since this is akin to a random walk. On the other
hand, multiscaling is observed both in RFM and in variation (i) models.
We notice that deviations from simple self-affine scaling 
are stronger for high $q$ values, while for low $q$ one recovers
the $\alpha_q = \zeta_{loc}$. The origin of this behavior is
related to the removal of overhangs on the crack surface, 
a process that inevitably produces steps in the 
single valued crack profile. As discussed by Mitchell, adding
random steps to a self-affine profile yields an apparent
multiscaling over small scales and for $q>1$, while for low $q$ values, 
one obtains $\alpha_q = \zeta_{loc}$ \cite{mitchell05}. 
Therefore, we conclude that
the apparent multiscaling in RFM and in variation (i) models at small scales and 
for $q>1$ is due to the process of removal of overhangs, and that
pure self-affinity is recovered at larger scales. This can be further
elaborated by considering the height difference $\Delta h(l) = |h(x+l)-h(x)|$.
We have checked its scaling with $l$ and varying $L$, and three conclusions
can be drawn: (1) $\langle \Delta h(l) \rangle \sim l^{\zeta_{loc}}$ as expected. 
(2) The $L$-scaling exhibits similar anomalous scaling as the width and
power spectrum data.  (3) The crack profile is stationary (e.g. $\langle \Delta h(1) \rangle$ does not depend on $x$). Hence, the observed anomalous scaling is not due to 
non-stationarity of the crack growth process but rather to an intrinsic dependence
of the crack thickness on $L$.

\begin{figure}[hbtp]
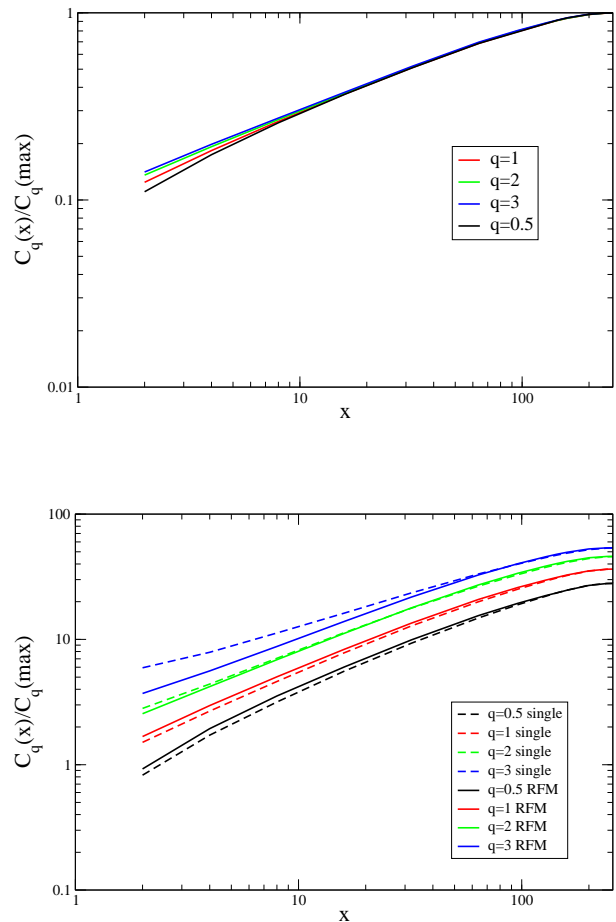

\begin{center}
\includegraphics[width=8cm]{directed.eps}
\end{center}
\vspace{0.5cm}
\begin{center}
\includegraphics[width=8cm]{compare.eps}
\end{center}
\caption{The $q$-correlation functions, normalized by their maximum
values, for directed cracks (top),  connected cracks and
RFM (bottom). Multiscaling is essentially absent for directed crack, while both connected craks and the RFM display deviations from
simple scaling, especially at high $q$. Notice that the multiscaling behavior of
connected cracks and RFM follows a very similar pattern.}
\label{fig:3}
\end{figure}

Finally Fig.~4 depicts the scaling of the histograms of $\Delta h(l)$ 
for $L=512$, and for variation (i).
Empirical analysis of 2d fracture data implies that for values of
$l$ such that
the profiles are self-affine, the distributions of $\Delta h$
often approach Gaussian 
(\cite{salminen03}, see also \cite{santucci06} for a similar,
claim). Here, we can discern clearly a Gaussian
central part, but the tails do not follow a Gaussian even for $l=64$
(similar to what is seen in Ref.~\cite{salminen03}). Varying
$L$ reveals again the presence of the anomalous scaling.

\begin{figure}[hbtp]
\begin{center}
\includegraphics[width=8cm]{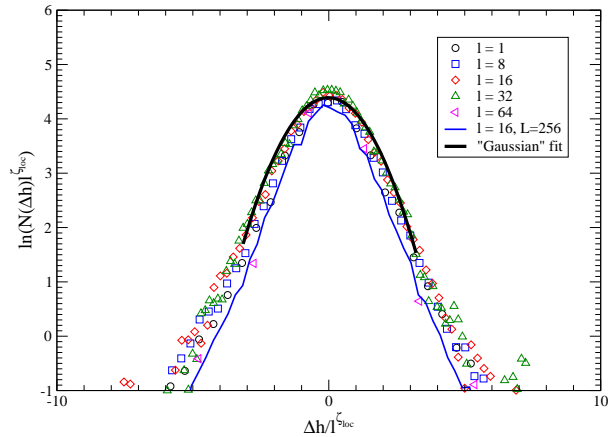}
\end{center}
\caption{The (logarithms) of histograms of $\Delta h(l)$ for $l = 1,8,16,32,64$ and $L=512$.
As a guide for the eye we present a Gaussian fit for $l=32$. It can be seen that the
central parts of the distributions are Gaussian, but for large $\Delta h$ deviations
exist, probably originating from the overhangs that are removed. In addition, we report
the data for $L=256$ and $l=16$ to show the systematic deviations due to the anomalous 
scaling with $L$}
\label{fig:4}
\end{figure}

In summary, having analyzed the crack morphology of two 
dimensional fracture surfaces in experiments and models,
we conclude that multiscaling is an
artifact due to the removal of small scale overhangs and
that self-affinity is recovered at large scales. The study of
single crack variations of the original RFM indicates that
the roughness properties of the fracture surface 
are due to the damage accumulation 
within the FPZ surrounding the crack, whereas the diffusive 
damage nucleation distributed homogeneously over the rest of 
the geometry is irrelevant.
Finally, a model of a moving crack tip appears to
be an oversimplifaction of the problem, and yields quantitatively 
different results from those of the RFM.
It is intriguing to speculate about the three dimensional
case, where crack front line models have enjoyed a wide
appeal in the literature but are always in quantitative
disagreement with experiments. Simulations of connected
cracks with large scale three dimensional fuse models 
could help clarify this issue.

\bibliography{../../biblio/fracture}
\bibliographystyle{iopart-num}

\end{document}